\documentclass[letterpaper,times]{IONconf}
\usepackage[pdftex]{graphicx}
\DeclareGraphicsExtensions{.pdf,.jpeg,.png}

\usepackage{amsmath}
\usepackage{amssymb}

\usepackage[ruled]{algorithm2e}
\usepackage{algpseudocode}

\usepackage{array}
\usepackage{textcomp}
\usepackage{makecell}
\usepackage{hhline}

\usepackage{multicol}
\usepackage{multirow}
\usepackage{booktabs}
\usepackage{threeparttable}

\usepackage{verbatim}

\usepackage{xcolor}

\usepackage{pifont}% http://ctan.org/pkg/pifont

\newcolumntype{P}[1]{>{\centering\arraybackslash}p{#1}}

\newcommand{\trp}[1]{{#1}^{\mathsf{T}}}

\newcommand{\thickhline}{
    \noalign {\ifnum 0=`}\fi \hrule height 1pt
    \futurelet \reserved@a \@xhline
}

\makeatletter
\def\env@dmatrix{\hskip -\arraycolsep
  \let\@ifnextchar\new@ifnextchar
  \extrarowheight=2ex
  \array{*\c@MaxMatrixCols{>{\displaystyle}c}}}

\makeatother

\DeclareCaptionLabelSeparator{custom}{ }
\DeclareCaptionFormat{custom}
{%
    \textbf{#1#2}\textit{\small #3}
}
\usepackage{url}
\usepackage{natbib}
\usepackage[hidelinks]{hyperref}
\title{Reducing Computational Complexity of Rigidity-Based UAV Trajectory Optimization for Real-Time Cooperative Target Localization}

\author{
    Halim~Lee and Jiwon~Seo, \textit{Yonsei~University}% <- this '%' removes a trailing whitespace
    }

\begin{document}

\maketitle

% biography section. The * indicates a section excluded from numbering.
\section*{Biography}

% Biographies are defined as follows:
% \biography{Author name}{author biography text}

\biography{Halim Lee}{is a Ph.D. candidate in the School of Integrated Technology, Yonsei University, Incheon, South Korea. She received the B.S. degree in Integrated Technology from Yonsei University. In 2020, she was a visiting graduate student at the University of California, Irvine, USA. Her research interests include urban navigation, navigation safety, and localization and tracking.}

\biography{Jiwon Seo}{is an Underwood Distinguished Professor at Yonsei University, where he is a professor in the School of Integrated Technology, Incheon, South Korea. 
He also serves as an Adjunct Professor in the Department of Convergence IT Engineering at Pohang University of Science and Technology (POSTECH), Pohang, South Korea.
His research interests include GNSS anti-jamming technologies, complementary PNT systems, and intelligent unmanned systems.
Dr. Seo is a member of the International Advisory Council of the Resilient Navigation and Timing Foundation, Alexandria, VA, USA, and the Advisory Committee on Defense of the Presidential Advisory Council on Science and Technology, South Korea.}

% The Abstract. The * indicates a section excluded from numbering.
\section*{Abstract}

Accurate and swift localization of the target is crucial in emergencies. However, accurate position data of a target mobile device, typically obtained from global navigation satellite systems (GNSS), cellular networks, or WiFi, may not always be accessible to first responders. For instance, 1) accuracy and availability can be limited in challenging signal reception environments, and 2) in regions where emergency location services are not mandatory, certain mobile devices may not transmit their location during emergencies.
As an alternative localization method, a network of unmanned aerial vehicles (UAVs) can be employed to passively locate targets by collecting radio frequency (RF) signal measurements, such as received signal strength (RSS). In these situations, UAV trajectories play a critical role in localization performance, influencing both accuracy and search time. Previous studies optimized UAV trajectories using the determinant of the Fisher information matrix (FIM), but its performance declines under unfavorable geometric conditions, such as when UAVs start from a single base, leading to position ambiguity.
To address this, our prior work introduced a rigidity-based approach, which improved the search time compared to FIM-based methods in our simulation case. However, the high computational cost of rigidity-based optimization, primarily due to singular value decomposition (SVD), limits its practicality. In this paper, we applied techniques to reduce computational complexity, including randomized SVD, smooth SVD, and vertex pruning. Simulations demonstrate that computational complexity can be reduced from $\mathcal{O}(l \times \max(m, n) \times \min(m, n)^2)$ to $\mathcal{O}(1)$, where $l$ is the number of iterations for solving the optimization problem, and $m$ and $n$ are the dimensions of the rigidity matrix.
Despite the significant reduction in computational cost, our investigation found no notable decrease in target localization performance, including both search time and root mean squared error (RMSE) of the position estimates. The proposed approach enables real-time UAV-based target localization in emergency scenarios.

% The introduction. Section numbering starts here.
\section{INTRODUCTION}

Emergency location services (ELSs) have been actively studied \citep{Arafat19:Localization, Elsawy17:Base, Ferreira17:Localization, Lee22:Evaluation, Moon24:HELPS, Zhang12:Inertial}.
%(하림) 여기에 다른 연구그룹 논문 몇 개 추가해주세요 
Recently, modern enhanced 911 (E911) features integrated into mobile phones, such as Android's ELS \citep{Malkos18:Emergency}, have enabled the automatic transmission of global positioning system (GPS), cellular, or WiFi-based position estimates to 911 dispatchers during emergencies.
However, the accuracy and availability of these positioning methods are often limited, either by the signal-receiving environment (e.g., urban or indoor areas where GPS signals are blocked \citep{Kim24:Performance, Kim23:Low, Kim23:Machine, Lee23:Seamless, Lee22:Urban}), signal coverage (e.g., deserts or mountains where cell towers are absent), radio interference (e.g., jamming or spoofing of GPS signals \citep{Park21:Single, Jia21:Ground, Lee22:SFOL, Kim22:First}, or ionospheric anomalies \citep{Lee17:Monitoring, Lee22:Optimal, Sun20:Performance, Sun21:Markov}). 
Additionally, in countries where E911 service is not mandatory, some devices may not automatically report their positions during emergencies. 
In such situations, a network of unmanned aerial vehicles (UAVs) can be utilized to locate the target cooperatively \citep{Uluskan20:Noncausal, Wang18:Performance, Lee23:Performance_Evaluation, Lee25:Rigidity}. 
The target's position can be calculated using measurements (e.g., time-of-arrival (TOA), angle-of-arrival (AOA), and received signal strength (RSS)) from the radio frequency (RF) signal transmitted by the target, which are captured by UAVs.

In UAV-enabled cooperative target localization, the trajectory of UAVs plays a crucial role in determining the performance of target localization, impacting both the localization accuracy and the time needed to achieve the desired level of accuracy. Previous studies \citep{Uluskan20:Noncausal, Wang18:Performance, Lee23:Performance_Evaluation} have primarily employed the determinant of Fisher Information Matrix (FIM) as an optimization metric for UAV trajectories. The Cramer-Rao Lower Bound (CRLB), which is the reciprocal of the Fisher information, provides a lower bound on the variance of any unbiased estimator. However, our recent work \citep{Lee23:Impact} has confirmed that the performance of FIM-based UAV trajectory optimization (i.e., waypoint optimization) diminishes when the initial geometric diversity of UAVs is unfavorable, such as when all UAVs commence missions from a single base. In such unfavorable UAV geometry scenarios, the position ambiguity problem can occur, in which multiple candidate solutions yield similar sensor measurements.

To address the position ambiguity problem, our prior work \citep{Lee25:Rigidity} employed the concept of rigidity, primarily studied in geometry and combinatorics. Our proposed approach optimizes the trajectories of UAVs to move in a direction that makes the graph comprising the UAV and target more rigid \citep{Lee25:Rigidity}. In our prior work, we demonstrated that our rigidity-based approach can enhance target localization performance in terms of search time compared to the existing FIM-based approach \citep{Lee25:Rigidity}.

Despite its commendable performance, rigidity-based UAV trajectory optimization suffers from a drawback due to its high computational cost, primarily attributed to the repetitive singular value decomposition (SVD) operation of the rigidity matrix. The computational cost of the rigidity-based approach is amplified for the following three reasons. Firstly, the SVD operation itself is costly, with MATLAB's ``svd'' function operating at $\mathcal{O}(\max(m,n) \times \min(m,n)^2)$ complexity for $m \times n$ matrix \citep{Wang15:Practical}. Secondly, due to the iterative nature of solving the optimization problem, the SVD needs to be repeatedly performed. Lastly, as the size of the graph comprising UAVs and the target expands over time, the computational cost of the SVD operation escalates accordingly.

To overcome these limitations, we developed a method to reduce the computational complexity of the rigidity-based UAV trajectory optimization approach in this paper. Specifically, we have applied three techniques to mitigate the aforementioned computational cost issues: 1) randomized SVD, 2) smooth SVD, and 3) vertex pruning  \citep{Lee25:Rigidity}. Among them, vertex pruning has been introduced in our prior work \citep{Lee25:Rigidity}. We verified the localization performance of the proposed method, which utilizes all three computational cost-mitigation techniques: randomized SVD, smooth SVD, and vertex pruning, through simulation.

\section{Problem Description} 
\label{sec:ProblemDescription}

\subsection{UAV-based emergency localization}
We consider an emergency localization scenario where $I$ UAVs are equipped with receivers capable of measuring RSS from a signal transmitted by the target. Each UAV is assumed to determine its own position using onboard navigation systems such utilizing GNSS \citep{Seo14:Future, Zhang18:Intelligent, Tang22:GNSS} or alternative navigation methods \citep{Kassas15:Greedy, Khalife22:Achievability, Kim14:Multi, Lee19:Safety, Rhee21:Enhanced, Yang21:UAV}. 

At any time $t \geq 0$, the position of the $i$-th UAV is denoted by $\mathbf{x}^\mathrm{UAV}_{i, t} = \trp{\left[ x^\mathrm{UAV}_{i, t}, y^\mathrm{UAV}_{i, t} \right]}$. Let $\mathbf{r} = \trp{\left[ x_\mathrm{r}, y_\mathrm{r} \right]}$ represent the true location of the target. The RSS measurement recorded by the $i$-th UAV at time $t$, labeled as $\hat{P}_{i, t}$, is modeled as follows \citep{Lee23:Performance_Evaluation}:
\begin{equation} \label{eq:Path_loss_model}
\begin{split}
\hat{P}_{i, t} &= P_{i,t} + N_{i, t} = P_{0} - 10 \beta \log_{10} \frac{s_{i, t}}{s_0} + N_{i, t},\\
s_{i, t} &= \|\mathbf{x}^\mathrm{UAV}_{i,t}-\mathbf{r}\|, \\
N_{i, t} &\sim \mathcal{N}(0, \sigma_\mathrm{dB}^2),
\end{split}
\end{equation}
where $P_{0}$ (in dBm) represents the RSS at a reference distance $s_0$ from the target, assumed to be 1 m in this work for simplicity. The path loss exponent is denoted by $\beta$, while $N_{i, t}$ models log-normal shadowing as a normal distribution in dB, with variance $\sigma_\mathrm{dB}^2$.

\subsection{Rigidity-based UAV trajectory optimization}
Our objective is to determine the optimal waypoint for the $i$-th UAV at time $t+1$ (i.e., $\mathbf{x}^\mathrm{UAV}_{i, t+1}$) that minimizes the target localization error as effectively as possible, based on the RSS measurements collected up to time $t$.

Previous studies \citep{Uluskan20:Noncausal, Wang18:Performance, Lee23:Performance_Evaluation} utilized the determinant of the FIM to guide UAVs in a way that minimizes the target localization error covariance. However, our prior work \citep{Lee25:Rigidity} showed that in scenarios with position ambiguity in target localization, a rigidity matrix-based metric is more effective at reducing target search time than using the determinant of the FIM.
Position ambiguity becomes more pronounced when RSS errors are significant and the UAV-target geometry is unfavorable.

According to our prior study \citep{Lee25:Rigidity}, the UAVs move in the direction that maximizes the $(d|V|-d(d+1)/2)$-th singular value of the rigidity matrix $R \in \mathbb{R}^{m \times n}$ of the body-bar framework formed by the UAVs and the target. Here, $d$ represents the dimension of the space ($d=2$ in this study as we consider a two-dimensional space), and $|V|$ denotes the total number of vertices, which includes all UAV positions untill time $t$ and the target. For more detailed information on constructing the rigidity matrix, please refer to our prior work \citep{Lee25:Rigidity}.

\section{Computational Complexity Reduction Methods} \label{sec:Methods}
To compute the singular value of the rigidity matrix, SVD operation is required. However, real-time optimization of UAV trajectories can be computationally demanding due to the heavy computational load of the SVD. As mentioned before, 1) the SVD operation itself (e.g., MATLAB's ``svd'' function) is costly, 2) the computation must be repeated for all candidate moving directions, and 3) as time progresses and the UAVs move, the size of the rigidity matrix $R$ increases.

To tackle these three challenges, we employed the following techniques: 1) randomized SVD, 2) smooth SVD, and 3) vertex pruning.

\subsection{Randomized SVD}

To reduce the computational cost of SVD, randomized SVD \citep{Halko11:Finding} is applied.
Randomized SVD is an efficient technique for approximating the SVD of a matrix, particularly useful for large-scale problems. Instead of directly computing the SVD of the original matrix $R \in \mathbb{R}^{m \times n}$, which can be computationally expensive, this method leverages random projections to approximate the range of $R$.

The procedure for randomized SVD is as follows.
First, a random matrix $\Omega \in \mathbb{R}^{n \times p}$ is generated, where $p = \min(2k, n)$ and $k$ represents the desired rank, defined in this paper as $k = 2|V| - 3$. The product $Y = R \cdot \Omega$ is then computed, with $Y \in \mathbb{R}^{m \times p}$ capturing the most significant features of $R$ in a lower-dimensional subspace. Subsequently, the columns of $Y$ are orthonormalized to form $Q \in \mathbb{R}^{m \times p}$, which approximates the column space of $R$. Once the matrix $Q$ is obtained, the original matrix $R$ is projected into this lower-dimensional subspace as $B = Q^\top \cdot R \in \mathbb{R}^{p \times n}$.

Then, the SVD of $B$ is computed as \citep{Halko11:Finding}:
\begin{equation}
B = \tilde{U} \cdot \Sigma \cdot V^\top,
\end{equation}
where $\tilde{U} \in \mathbb{R}^{p \times k}$, $\Sigma \in \mathbb{R}^{k \times k}$, and $V \in \mathbb{R}^{n \times k}$.

Finally, the left singular vectors of $R$ are reconstructed as $U = Q \cdot \tilde{U}$. The result is a low-rank approximation of $R$ \citep{Halko11:Finding}:
\begin{equation}
R \approx U \cdot \Sigma \cdot V^\top,
\end{equation}
where $U \in \mathbb{R}^{m \times k}$, $\Sigma \in \mathbb{R}^{k \times k}$, and $V \in \mathbb{R}^{n \times k}$.

Randomized SVD has a computational complexity of $\mathcal{O}(mnk)$ for large matrices, where $k \ll \min(m, n)$ and $p$ is small. This is significantly lower than the $\mathcal{O}(mn^2)$ or $\mathcal{O}(m^2n)$ complexity of traditional SVD methods.

\subsection{Smooth SVD}

To reduce the computational cost of repeatedly performing SVD for all candidate moving angles, smooth SVD \citep{Dieci99:Smooth} is applied. Smooth SVD is a method of updating singular values for small perturbated or smooth time-varying matrices \citep{Dieci99:Smooth}. When smooth SVD is applied, the differential of the singular values can be expressed as a function of the initial SVD and the differential of the rigidity matrix. Consequently, the subsequent SVD can be approximated through simple matrix multiplication once the initial SVD has been computed.

Let the initial singular value decomposition $R(\alpha_0) = U(\alpha_0) \Sigma(\alpha_0) V^*(\alpha_0)$ be given, where $\alpha_0$ is the initial moving angle. The matrix $R(\alpha)$, which depends on the candidate moving angle $\alpha$, can be expressed as: 
\begin{equation}
R(\alpha) = U(\alpha) \Sigma(\alpha) V^*(\alpha).
\end{equation}
Here, $U(\alpha)$, $\Sigma(\alpha)$, and $V(\alpha)$ represent the angle-dependent left singular vectors, singular values, and right singular vectors, respectively.

From the decomposition $R = U \Sigma V^*$, the derivative of $R$ yields the relation \citep{Dieci99:Smooth}:
\begin{equation}
U^* \dot{R} V = (U^* \dot{U}) \Sigma + \dot{\Sigma} + \Sigma (\dot{V} * V).
\end{equation}
Let $H := U^* \dot{U} \in \mathbb{C}^{m \times m}$ and $L := V^* \dot{V} \in \mathbb{C}^{n \times n}$, both of which are skew-Hermitian matrices (i.e., $H^* = -H$ and $L^* = -L$). When the singular values of $R(\alpha)$ are distinct, the diagonal elements $H_{jj}$ and $L_{jj}$ vanish. This simplifies the computation of the derivative of the singular values $\sigma_j(\alpha)$ as:
\begin{equation}
\dot{\sigma}_j(\alpha) = (U(\alpha_0)^* \dot{R}(\alpha) V(\alpha_0))_{jj},
\end{equation}
where $U_0$ and $V_0$ denote the initial left and right singular vector matrices.

Thus, by knowing the initial singular value and the difference of $R$ (i.e., $R(\alpha) - R(\alpha_0)$), we can approximate the singular value of $R(\alpha)$ for any given $\alpha$.

\subsection{Vertex Pruning}
To prevent the increase in computational cost as the size of the rigidity matrix grows, vertex pruning \citep{Lee25:Rigidity} is applied. Vertex pruning is a method of eliminating vertices that are deemed less informative as the size of the graph increases beyond a certain threshold \citep{Lee25:Rigidity}. When the number of UAV vertices to retain is $K$, the priority queue keeps only the $K$ highest-priority elements, removing all others. For further details, see \citet{Lee25:Rigidity}.

\section{Evaluation} \label{sec:Evaluation}
\subsection{Simulation Setting} \label{subsec:SimulationSetting}

In the simulation, it is assumed that the target is located at (0 m, 0 m), and two UAVs start the mission at (-125 m, -125 m) and (-125 m, -122.5 m). RSS errors are assumed to follow a normal distribution with a mean of 0 and a standard deviation of 5 dB, in accordance with the ITU-R recommendation for Suburban Macro or Rural Macro Scenarios \citep{ITUR09}. Additionally, the signal transmission power of the target is assumed to be 3 dBm, and the path loss exponent is assumed to be 2. It is assumed that UAVs fly at 5 m/s and RSS can be measured every 1 s. To prevent sharp turns, the moving direction of the UAV is restricted to within $\pm$20 degrees of the moving direction from the previous epoch. The optimal moving direction is determined through a brute-force search, wherein the objective value is calculated for all angles at 1-degree intervals within the 40-degree range. Simulations were performed 250 times for each method. All code was written in MATLAB, and all simulations were executed on an Intel(R) Xeon(R) CPU E5-2699 operating at 2.30 GHz.

\subsection{Simulation Result} \label{subsec:SimulationResult}

\begin{figure}[htp!]
    \centering
    \includegraphics[width=\linewidth]{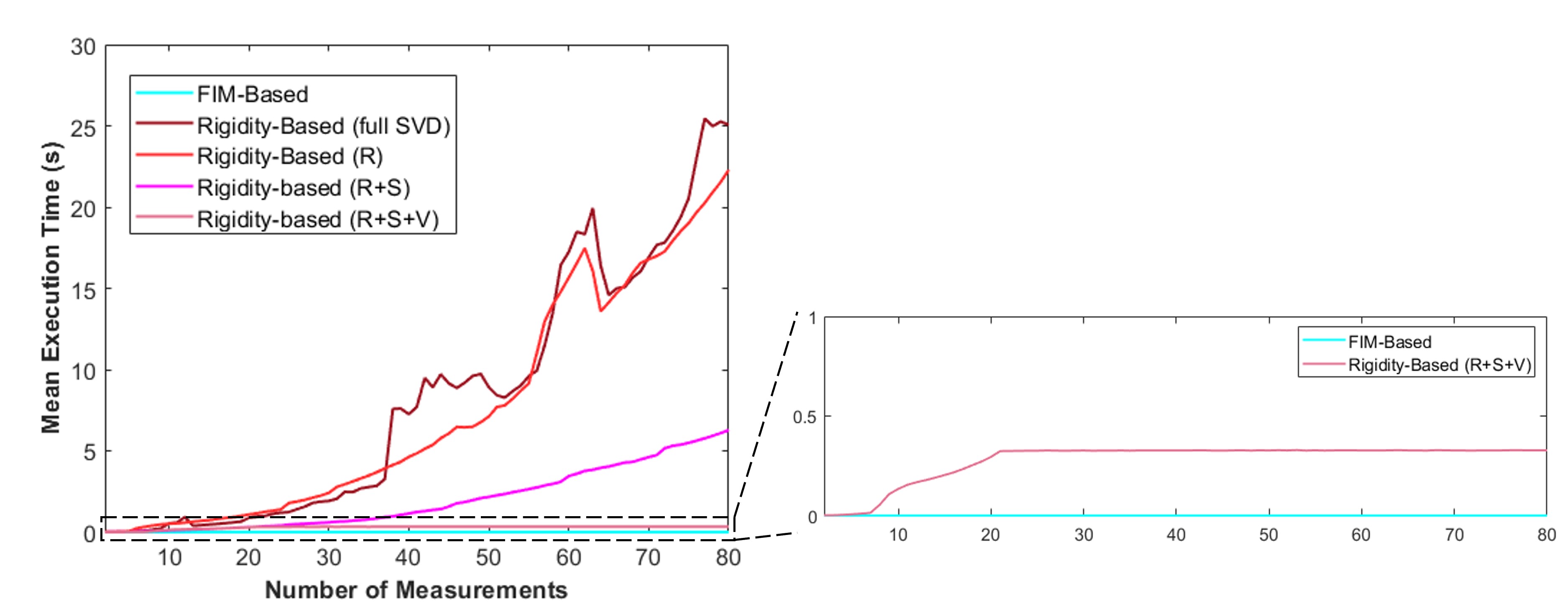}
    \caption{The mean code execution time as measurements accumulate.}
    \label{fig:MeanExecutionTime}
\end{figure}

Firstly, we verified the mean code execution time of our proposed method. Fig. \ref{fig:MeanExecutionTime} shows the mean code execution time of each case. Fig. \ref{fig:MeanExecutionTime} presents the execution time for solving the UAV trajectory optimization problem, excluding other operations such as position estimation. `Full SVD' refers to the case where the MATLAB’s ``svd'' function is used for all SVD operations without employing the computational cost reduction method proposed in this paper. `R' denotes the application of only randomized SVD. `R+S' indicates the use of both randomized SVD and smooth SVD. `R+S+V' indicates the proposed method, which uses randomized SVD, smooth SVD, and vertex pruning together.

Table \ref{tab:ComputationalComplexity} compares the computational complexity of solving the UAV trajectory optimization problem. In Table \ref{tab:ComputationalComplexity}, $l$ represents the number of iterations for optimizing the next moving angle of the UAV (e.g., $l=41$ when considering next possible moving angles at 1-degree intervals within the 40-degree range). $K$ is the number of measurements to retain when applying vertex pruning. The proposed `R+S+V' case maintains a constant computational complexity (i.e., $\mathcal{O}(1)$) in the long-term, as demonstrated in Fig. \ref{tab:ComputationalComplexity}, owing to vertex pruning which prevents the rigidity matrix from expanding beyond a certain size.

\begin{figure}[htp!]
    \centering
    \includegraphics[width=\linewidth]{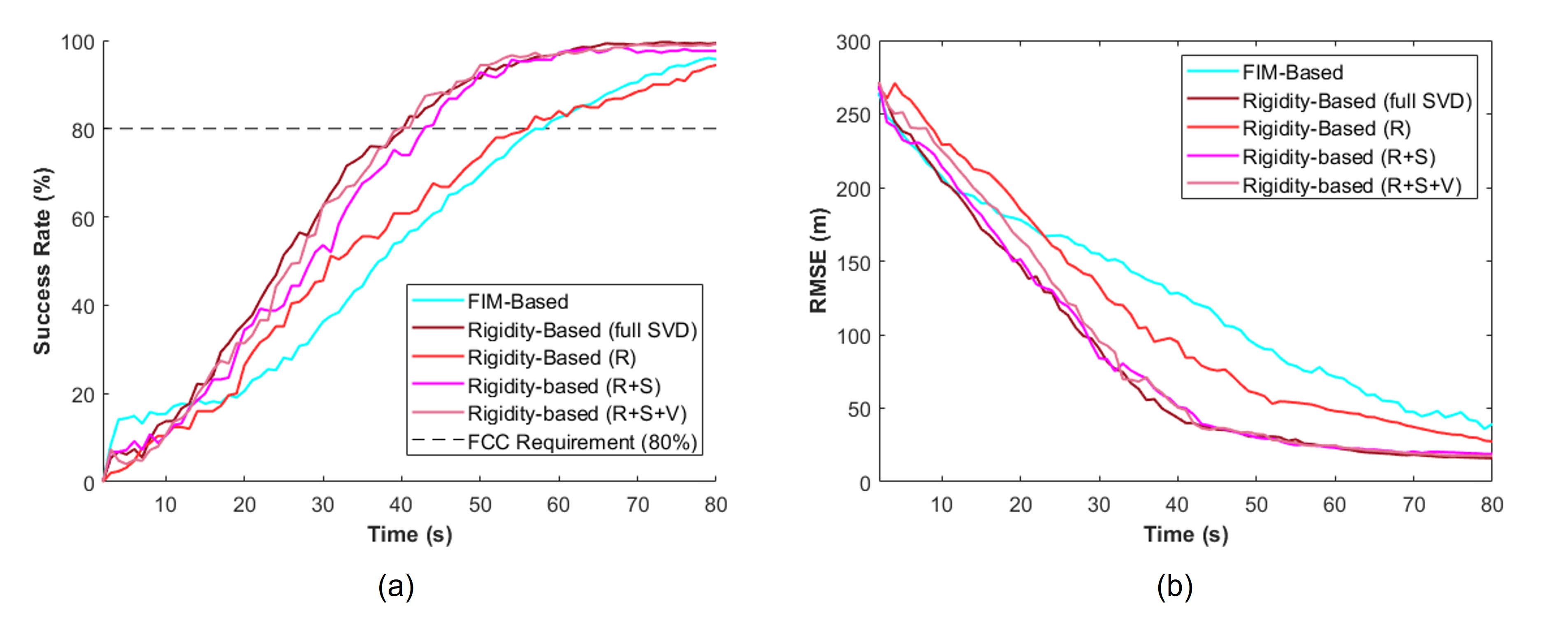}
    \caption{Target localization performance. (a) Success rate and (b) RMSE over time.}
    \label{fig:SimResults}
\end{figure}

Figs. \ref{fig:SimResults}(a) and (b) show success rate and localization root mean squared error (RMSE) of each approach over time. The success rate refers to the ratio of simulation instances that meet the FCC E911 requirement among the total simulation instances. In an emergency, the FCC mandates that more than 80\% of all 911 calls must be estimated with a position accuracy of 50 meters or less \citep{FCC19}. Compared to the existing method (i.e., `full SVD'), there was no noticeable performance reduction in the other cases except for the randomized SVD-only (i.e., `R') case.

\begin{table}[htp!]
\centering
\caption{Comparison of Computational Complexity}\label{tab:ComputationalComplexity}
\begin{center}{
\renewcommand{\arraystretch}{1.4}
 \begin{tabular}{ >{\centering\arraybackslash}p{3cm} | >{\centering\arraybackslash}p{11cm} }
 \hhline{==}
 \rule{0pt}{15pt} \thead{Operation} & \thead{Complexity} \\[1pt]
 \hline
 \rule{0pt}{15pt} Full SVD & $\mathcal{O}(l\times\max(m,n)\times\min(m,n)^2)$\\[4pt]
 \hline
 \rule{0pt}{15pt} R & $\mathcal{O}(l\times m\times n\times k)$\\[4pt]
 \hline
 \rule{0pt}{15pt} R+S & $\mathcal{O}(m\times n\times k)$\\[4pt]
 \hline
 \rule{0pt}{15pt} R+S+V (proposed) & \parbox[c][1.3cm][c]{11cm}{\centering When the number of measurements is less than $K$: $\mathcal{O}(m\times n\times k)$ \\ otherwise: $\mathcal{O}(1)$} \\
 \hhline{==}
\end{tabular}}
\end{center}
\end{table}

\section{Conclusion} \label{sec:conclusion}

To reduce the computational complexity of the rigidity-based UAV trajectory optimization method, we employed three techniques: 1) randomized SVD, 2) smooth SVD, and 3) vertex pruning. Our proposed method (i.e., `R+S+V’ case) has effectively minimized the computational complexity of solving the rigidity-based UAV trajectory optimization problem from $\mathcal{O}(l\times\max(m,n)\times\min(m,n)^2)$ to $\mathcal{O}(1)$. Despite achieving a significant reduction in computational cost, our investigation revealed no noteworthy decrease in target localization performance, including both search time and RMSE. Consequently, the computational cost reduction method proposed in this paper enables real-time cooperative target localization in emergency scenarios.

\section*{ACKNOWLEDGEMENTS}

This work was supported in part by the National Research Foundation of Korea (NRF), funded by the Korean government (Ministry of Science and ICT, MSIT), under Grant RS-2024-00358298; 
in part by the Future Space Navigation and Satellite Research Center through the NRF, funded by the MSIT, Republic of Korea, under Grant 2022M1A3C2074404; 
in part by Grant RS-2024-00407003 from the ``Development of Advanced Technology for Terrestrial Radionavigation System'' project, funded by the Ministry of Oceans and Fisheries, Republic of Korea;
in part by the Unmanned Vehicles Core Technology Research and Development Program through the NRF and the Unmanned Vehicle Advanced Research Center (UVARC) funded by the MSIT, Republic of Korea, under Grant 2020M3C1C1A01086407;
and in part by the MSIT, Korea, under the Information Technology Research Center (ITRC) support program supervised by the Institute of Information \& Communications Technology Planning \& Evaluation (IITP) under Grant IITP-2024-RS-2024-00437494.

% the apacite bibliography style matches the ION bibliography style guidelines.
\bibliographystyle{apalike}
\bibliography{mybibfile, IUS_publications}

\end{document}